# Space and Time as a Primary Classification Criterion for Information Retrieval in Distributed Social Networking

Georg Groh[1], Florian Straub[2], Andreas Donaubauer[2], Benjamin Koster[1]

[1]Chair for Applied Informatics / Cooperative Systems
Technische Universitaet Muenchen
Boltzmannstr. 3; 85748 Garching
Email: {grohg,straub,koster}@in.tum.de

[3]Institut für Geodäsie und Photogrammetrie
Eidgenössische Technische Hochschule Zürich
Wolfgang-Pauli-Str. 15; 8093 Zürich
Email: donaubauer@geod.baug.ethz.ch

## 1. Introduction

Decentralized Social Networking (DSN) is a key paradigm of the future Social Web (Yeung 2009), which is able to solve key interoperability problems ("Data Silo Problem"). Furthermore, mobile computing which blurs the boundaries between real and virtual world, is also strongly growing in importance for the Social Web (Fox 2010). The related stimulated usage of spatial referencing of information items and users in services, which is part of the increased spatialization of the Web, delivers an added filtering possibility for information. Furthermore, the aggravation of the Hidden Web problem (He 2007) caused by individual access control necessary to implement privacy with respect to the increasingly personal content is becoming a problem in view of effective IR in the Social Web. In this Participation-Web of prosumers, asking persons from the social network neighbourhood is becoming an increasingly important alternative model of information retrieval (IR) (Gibs 2009), implementing a decentralized autonomous natural human way of IR (Groh 2010). In this contribution we discuss the evaluation of key concepts of an architecture for an alternative, decentralized System of IR-agents, implementing these observations as described in Groh (2010).

## 2. Human IR, Small Worlds and Tobler's First Law

For a alternative IR approach it is to reasonable employ the mentioned Social Web IR mode of asking other users (cf. natural human principle of IR). Human IR and systems that emulate this type of IR will in general not be able to deliver precision and recall of data-retrieval systems (like databases) or conventional IR systems but will deliver information items and relations between such items that have not been reflected in the initial query or index structures but that (e.g. via social or spatiotemporal correlations) are nevertheless highly useful. The selection of the right persons in human IR is based on social, semantic, and spatiotemporal aspects, using existing small world structures and respective concepts of distance for routing in such networks (Kleinberg 2000). A connection between such decentralized personal search in small worlds and spatiotemporal reference is established via the so called Tobler's First Law of Geography (TFL) (Tobler 1970).

# 3. An Concept for Alternative Multi-Agent-System Based IR

For our architecture, employing human IR means asking the right agent as an implementation of asking the right person.

## 3.1 Spatio-Temporal Reference as Primary Classification Criterion

Applying TFL to graphs modeling a social or semantic network by locating the nodes in space implies the implicit conservation of the structural social or semantic small world properties of these graphs (Groh 2010). According to TFL, entities with small geographic distance are also statistically more socially and semantically related. We assume that analogous reasoning can be made for temporal references too. Together this gives rise to the notion of implicitly social (semantic) spatio-temporal small worlds. (Compare the schematic links (b) and (c) in figure (1)). The link (a) of figure (1), symbolizes connections between social distance and semantic distance (e.g. between documents in the information spaces of the respective persons) (see e.g. (Groh 2007)).

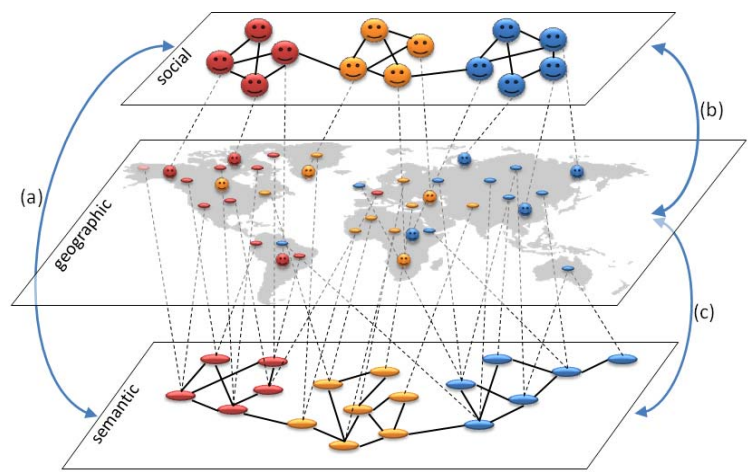

Figure 1. Connections between Social and Semantic Small Worlds and Spatio-Temporal Referencing.

In contrast to other approaches which use explicit social and semantic clustering for constructing coarse grained index structures or small world networks for IR, we focus on the spatio-temporal referencing of persons and information implicitly conserving social and semantic structures.

In our Multi-Agent-System (MAS), each agent maintains a number of information items which it wants to provide to the public by summarizing them in expertise profiles using spatio-temporal clustering. Furthermore it maintains a list of also spatio-temporally referenced Expert-Links to other agents as a means of estimating their expertise. Both are published into a global knowledge base (a common distributed spatio temporal index (DSTI) which has the right degree of granularity (Groh 2010)).

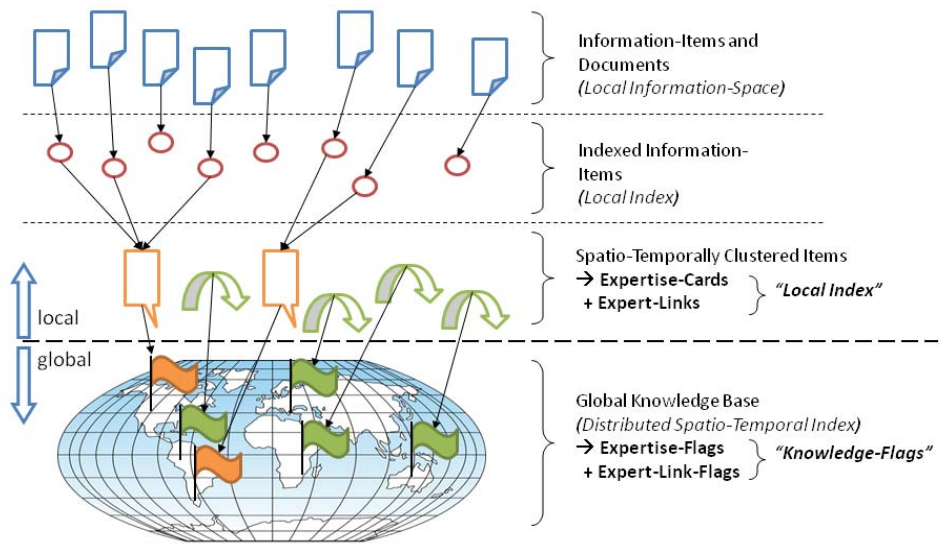

Figure 2. From Local Information Space to Published Knowledge-Flags via Spatio-Temporal Referencing (Groh 2010).

## 3.2 A Spatio-Temporal P2P Protocol

In the style of other P2P approaches, peers keep parts of the distributed index, which is spatio-temporally structured and based on a Quadtree approach. We use a distributed variant of Quadtree because in contrast to other approaches like CAN it can handle polygons, lines and other more expressive geometries. Furthermore, managing stabilization, splitting and merging can be done with less messages between peers. In contrast to the Distributed Quadtree (Tanin 2007) our approach is not build on top of an existing DHT as an application. Instead, the structure of the network follows the structure of the Quadtree.

With an increasing amount of added Expertise-Flags Expert-Link-Flags and Agent-Location-Flags the maximum capacity of a node (defined by the managing peer) is exceeded and splitting of a node is required. Finding the at most four peers residing in the child sectors can be achieved with local knowledge only, because peers insert their position and contact information to the tree. After an acknowledgement, mutual pointers between child and parent peer are added and content covered by the sector of a child is moved to the corresponding peer.

Routing in a P2P-Quadtree is based on the path of the node with the smallest sector that the geometry (e.g. of a query or insert) is covered by and the path of the node that a peer is either located in or managing.

## 4. Evaluation

A prototype of our P2P protocol and a simulation system was build, allowing to model influence, consequences and impact of parameters and paradigms. To simulate our approach with a realistic and comprehensive data set, we used the articles of the German and English Wikipedia with an explicit spatio-temporal reference. We accessed this dataset via the API provided by Hecht (2010). We added over 13,000 geometries of countries, $1^{st}$, $2^{nd}$ and $3^{rd}$ level administrative districts and nearly 30,000 maximum bounding boxes. The dataset was randomly distributed to 2000 virtual agents (peers) in a non-disjoint way. Several Expertises per agent were then computed by spatial clustering. Expert-Links were simulated by building a Small World with the help of a Kleinberg's Small World generator based algorithm (Kleinberg 2000). Our prototype then allows for queries to be issued allowing for an evaluation of the approach.

The evaluation showed that the designed P2P protocol using a distributed Quadtree as distributed spatio-temporal index fulfills the requirements that we derived from our concept for an alternative MAS for IR. A qualitative evaluation of the IR results showed that the approach is able to efficiently exploit the relations between spatio-temporal references and implicit social and semantic small world structures. For queries without spatio-temporal reference the qualitative evaluation showed that useful results were delivered by beginning the search in the spatio-temporal vicinity of the current location of the querying user.

## 5. Future Work

Our main focus for future work is the development of an appropriate performance measure which is able to quantify the usefulness of IR results in relation to an information need instead of modeling rather strict measures of relevance (such as precision and recall). Answers delivered by our approach that do not precisely match the query can nevertheless useful for the problem or information need leading to the query, because of implicit social and semantic small world effects mediated by spatio-temporal reference. By replacing immediate relevance to a query with a model of usefulness, we will be able to quantitatively characterize the performance of our alternative approach and to further improve key aspects of our architecture.